\documentclass{JHEP} \newcommand{\be}{\begin{equation}} 
\newcommand{\ee}{\end{equation}} \newcommand{\rf}[1]{(\ref{eq:#1})} 
\newcommand{\X}{{\cal X}} \newcommand{\Z}{{\cal Z}}

\newcommand{\M}{ {\cal M}}

\title{Brane-world Quantum Gravity}

\author{M.D. Maia\footnote{maia@unb.br}, Nildsen Silva \footnote{nildsen@hotmail.com} and M.C.B. Fernandes\footnote{mcezar@fis.unb.br}\\
Instituto de F\'\i sica, Universidade de Bras\'\i lia,\\
 70919-970, Bras\'\i lia, D. F., Brazil}


\abstract{The Arnowitt-Deser-Misner canonical formulation of general 
relativity is extended to the covariant brane-world theory in 
arbitrary dimensions. The exclusive probing of the extra dimensions 
makes a substantial difference, allowing for the construction of a 
non-constrained canonical theory. The quantum states of the 
brane-world geometry are defined by the Tomonaga-Schwinger equation, 
whose integrability conditions are determined by the classical 
perturbations of submanifolds contained in the Nash's differentiable 
embedding theorem. In principle, quantum brane-world theory can be 
tested by current experiments in astrophysics and by near future 
laboratory experiments at Tev energy. The implications to the 
black-hole information loss problem, to the accelerating cosmology, 
and to a quantum mathematical theory of four-sub manifolds are 
briefly commented.}

\begin{document}

\maketitle

\section{Quantizing the Brane-World} \vspace{3mm} 
\phantom{x}\hspace{4cm}\begin{minipage}{11cm}
 \emph{If gravity is to occupy a significant place
  in modern physics, it can do so only by being qualitatively
   different from other fields. As soon as we assume that
    gravity behaves qualitatively like other fields,
    we find that it is quantitatively insignificant}
  \phantom{x} \hspace{12cm} C.W. Misner (1957)
 \end{minipage}
 \vspace{3mm}

After analyzing perturbative quantum gravity, Misner reached the 
interesting conclusion that an effective quantum gravity must have 
qualities which makes it different from gauge theories 
\cite{Misner}. Translating quantitative significance in terms of 
energy level, Misner's conclusion suggests that the problem of 
quantization of the gravitational field should be solved 
concomitantly with the hierarchy problem of the fundamental 
interactions. In what follows, we apply this criterium to 
brane-world gravity.

Brane-world gravity is based on a higher dimensional solution of the 
hierarchy problem. In a seminal paper N. Arkani-Hamed, G. Dvali and 
S. Dimopolous questioned the currently accepted hypothesis that 
gravitons are quantitatively relevant only at the Planck scale of 
energies, essentially because this is an assumption devoid of 
experimental support. They proposed that the known gauge fields (and 
hence all ordinary matter) are to be confined within the 
four-dimensional brane-world, but gravitons can propagate in a 
higher-dimensional space, the bulk, at the same Tev scale of 
energies of the gauge fields \cite{ADD} (For historical papers on 
the development of the theory see also 
\cite{Antoniadis,Akama,Rubakov,RT,Visser,HW}). According to this 
view, brane-world gravity is qualitatively distinct from, but it is 
quantitatively equivalent to the gauge fields of the standard model.

Brane-world gravity predicts the existence of short lived Tev mini 
black holes, which in principle can be produced at the laboratory by 
a high energy proton-proton collision, with implications to the 
black hole information loss problem at the quantum level. The 
proposed experiment is set in Minkowski space-time, but it ends in a 
Schwarzschild (or Reissner-Nordstrom) space-time \cite{Landsberg}. 
Therefore, the theory supporting this experiment must be compatible 
with cross sections of the order of the Schwarzschild radius, and 
also with an explanation on how the original Minkowski space-time 
deforms into black hole, and back in a short period of time.

 Brane-world gravity may also explain the acceleration of the universe (see eg \cite{MaiaGDE} and references therein). In short, due to the presence of the extrinsic curvature, the vacuum in brane-world gravity is richer than the vacuum in general relativity. Besides the cosmological constant, it also contain a conserved geometric tensor built from the extrinsic curvature. Consequently, when studying the quantum fluctuations of such vacuum
we may obtain a different estimate for the vacuum energy density as 
compared with the case of general relativity.

Since most of the current research on brane-world theory is based on 
models defined in a five-dimensional bulk, using specific 
coordinates and particular symmetries (see eg \cite{RS}), we find it 
necessary to review in the next section the covariant equations of 
motion of a brane-world defined in an arbitrary bulk, with an 
arbitrary number of dimensions. Those equations can be found 
elsewhere \cite{Maeda,MaiaGDE}, but here we have included some 
details which are required for the quantum description. Readers who 
are familiar with this may jump to section 3, where the canonical 
equations of the brane-world with respect to the extra dimensions 
are discussed. In section 4 we introduce the Tomonaga-Schwinger 
equation for the brane-world with respect to the extra dimensions 
and comment on its integrability.

\section{Covariant Brane-world Gravity} There are essentially three 
basic postulates in brane-world theory: (1) The bulk geometry is 
defined by Einstein's equations; (2) The brane-world is a sub 
manifold embedded in that bulk ; (3) The gauge fields and ordinary 
matter are confined to four dimensions, but gravitons propagates 
along the extra dimensions at Tev energy \cite{ADD}.

 The embedding of the brane-world in the bulk plays an essential role on the covariant (that is, model independent) formulation of the brane-world gravity, because it tells how the Einstein-Hilbert dynamics of the bulk is transferred to the brane-world. However, there a are many different ways to embed a manifold into another, classified as local, global, isometric, conformal (or more generally defined by a collineation), rigid, deformable,
analytic or differentiable. The choice of one or another depend on 
what the embedded manifold is supposed to do.

In string theory the action principle is defined on the 
world-sheets, with additional boundary conditions, so that the 
embedding is necessarily global. Since the world sheets are 
2-dimensional they are all conformally flat and their global 
embedding is not difficult to achieve. However, if 
higher-dimensional objects such as p-branes are to be considered, 
then the global embedding may turn out to be difficult to realize in 
10 or even in 11 dimensions \cite{Friedman}.

Differently from string theory, the Einstein-Hilbert action in 
brane-world theory is set on the bulk, which is therefore the 
primary dynamical object. Furthermore, the embedding is locally 
defined, meaning that the bulk is a local fiber bundle whose fibers 
are the direct sum of the tangent and normal spaces at each point of 
the brane-world taken as the base space. If we want to draw a 
picture, the bulk can be seen as as a locally constructed space 
around each point of the brane-world.

A local differentiable embedding requires only that the embedding 
functions are differentiable and regular. This follows from Nash's 
embedding theorem, an important improvement over the traditional 
analytic embedding theorems of Janet and Cartan \cite{Janet,Cartan}, 
which demand that the embedding functions are represented by 
convergent positive power series. Furthermore, Nash's theorem shows 
that any sub manifold can be generated by a continuous sequence of 
small perturbations of an arbitrarily given sub manifold 
\footnote{The perturbative approach to the embedding was originally 
proposed by J. E. Campbell in 1926. However, his result differs from 
Nash's theorem because analytic conditions where implicitly used 
\cite{Campbell,Dahia}. Since the perturbation procedure is based on 
regular and differentiable functions, the differentiable embedding 
is less restrictive to the geometry than the analytic embeddings.}. 
Although the theorem was originally demonstrated for the case of an 
Euclidean bulk, it was later generalized to pseudo Riemannian 
manifolds \cite{Nash,Greene}.

Given a particular Riemannian sub-manifold $\bar{\sigma_4}$, its 
local isometric embedding in a certain bulk $\M_D$, is given by 
$D=4+N$ differentiable and regular embedding maps $\bar{\cal X}^A: 
\bar{\sigma}_4 \rightarrow \M_D$, such that \footnote{ Capital Latin 
indices refer to the bulk, which is a Riemannian geometry with 
metric
 ${\cal G}_{AB}$ in arbitrary coordinates. Small case Latin indices refer to the extra
 dimensions going from 5 to D, and all Greek indices refer to the brane, from 1 to 4. A curly $\cal
 R$ always denotes bulk curvatures, like in ${\cal R}_{ABCD}$.
 Ordinary capital $R$ like in
 $R_{\mu\nu}$ denotes brane-world curvatures.
Covariant derivatives need to be specified, for the bulk or the 
brane-world metrics. For a vector $V^A$ in the bulk its covariant 
derivative with respect to ${\cal G}_{AB}$ is denoted as $V^A_{;B}$. 
On the other hand, from the point of view of the brane-world metric, 
the components $V^A$ behave as a set of $N$ scalar functions as in
 \cite{Eisenhart}. For generality we denote ${\cal
 G}=|\mbox{det}( {\cal G}_{AB})|$.}
\be \bar{\X}^{A}_{,\mu}\bar{\X}^{B}_{,\nu}{\cal G}_{AB} 
=\bar{g}_{\mu\nu},\;\; \bar{\X}^{A}_{,\mu}\bar{\eta}^{B}_{b}{\cal 
G}_{AB}=0,\; \mbox{and} 
\;\;\bar{\eta}^{A}_{a}\bar{\eta}^{B}_{b}{\cal G}_{AB}=\bar{g}_{ab} 
\label{eq:X} \ee
 where $\bar{\eta}^A_a$ are the components of the $N$ linearly independent vector fields in the same coordinates of the bulk where the components ${\cal G}_{AB}
$ of the bulk metric are defined. The vectors $\{ \bar{\X}^A_{,\mu}, 
\bar{\eta}^B_a\}$ define a Gaussian reference frame called here the 
embedding frame. The derivatives of the vectors $\bar{\eta}_a$ is 
expressed in terms of the second and third fundamental forms 
$\bar{k}_{\mu\nu a}$, $\bar{A}_{\mu a b}$ respectively by the 
Gauss-Weingerten equations \cite{Eisenhart} \be 
\bar{\eta}^A_{a,{}\alpha} = \bar{g}^{\mu\nu}\bar{k}_{\alpha\mu 
a}\bar{\X}^A_{,\nu} +\bar{g}^{mn}\bar{A}_{\alpha a m }\bar{\eta}^A_n 
\label{eq:GW} \ee Without loss of generality we may chose the normal 
vectors $\bar{\eta}_a$ to be orthogonal to each other, so that 
$\bar{g}_{ab}= \epsilon_a \delta_{ab}$, where $\epsilon_a=\pm 1$ 
depending on the signature of the bulk \cite{Greene}.

 Nash's perturbative approach to embedding consists in subjecting the
fundamental forms of $\bar{\sigma}_4$ to small parametric deviations 
along each normal vector. It can be also described by introducing a 
small perturbation with parameter $ \delta y^a$, of the base vectors 
$\{{\bar \X}^A_{,\mu}, \eta^A_a\}$ along each normal 
$\bar{\eta}^{A}_{a}$ evaluated on $\bar{\sigma}_4$, obtaining 
another set of vectors (no sum on a) \begin{eqnarray}
\Z^{A}_{,\mu} &=&\bar{\X}^{A}_{,\mu} +(\delta y^a\pounds_{\bar{\eta}_a}\bar{\X}^{A})_{,\mu} =\bar{\X}^{A}_{,\mu} -\delta y^a[\bar{\X},\bar{\eta}_a]^A_{,\mu}=\bar{\X}^{A}_{,\mu} +\delta y^a\bar{\eta}^A_{a,\mu} ,\label{eq:pert} \\
\eta^{A}_a &= &\bar{\eta}^{A}_a + (\delta 
y^a\pounds_{{\bar{\eta}}_a}\bar{\eta}_a)^{A} =\bar{\eta}^{A}_a + 
\delta y^a[\bar{\eta}_a, \bar{\eta}_a]^{A} =\bar{\eta}_a^{A} 
\end{eqnarray} which define a perturbed embedding frame $\{ {\cal 
Z}^{A}_{,\mu}, {\eta}^{A}_{a}\}$ in the bulk. Admitting that these 
new functions remain differentiable and regular and that they 
satisfy the equations similar to \rf{X}, \begin{equation} 
\phantom{x}\hspace{-0.5cm}{\cal Z}^{A}_{,\mu}{\cal 
Z}^{B}_{,\nu}{\cal G}_{AB} =g_{\mu\nu},\;\; {\cal 
Z}^{A}_{,\mu}{\eta}^{B}_{a}{\cal G}_{AB}=g_{\mu a},\;\; 
{\eta}^{A}_{a}{\eta}^{B}_{b}{\cal G}_{AB}=g_{ab}=\bar{g}_{ab} 
\label{eq:Z} \end{equation}
 we obtain a $N$-parameter local family of submanifolds $\sigma_4$ generated by local perturbations of $\bar{\sigma}_4$, by a continuous variations of the parameters $\delta y^a$.

The next problem is to find a solution of these equations. However, 
instead of finding the coordinates $\Z^A$, it is more convenient to 
write the perturbed solution in terms of the fundamental forms, 
expressed in terms of the initial geometry of $\bar{\sigma_4}$. By 
direct substitution of ${\mathcal{Z}}_{,\mu}$ and $\eta^A_a$ derived 
from \rf{pert} in equations \rf{Z} we obtain \begin{eqnarray} 
g_{\mu\nu}(x,y) & = & 
{\mathcal{Z}}_{,\mu}^{A}{\mathcal{Z}}_{,\nu}^{B}{\mathcal{G}}_{\!
AB}=\bar{g}_{\mu\nu}\!\!- 2\delta y^{a}\bar{k}_{\mu\nu a} \nonumber \vspace{1mm}\\
\hspace{-8mm} &+&\delta y^{a}\delta 
y^{b}[\bar{g}^{\alpha\beta}\bar{k}_{\mu\alpha a}\bar{k}_{\nu\beta 
b}+g^{cd}\bar{A}_{\mu
ca}\bar{A}_{\nu db}],\label{eq:gmunu}\vspace{1mm}\\
g_{\mu a}(x,y) & = &
{\mathcal{Z}}_{,\mu}^{A}{\eta}_{a}^{B}{\mathcal{G}}_{\! AB}=\!\! \delta y^{b}A_{\mu ab}, \label{eq:gmua}\vspace{1mm}\\
g_{ab}(x,y) & = &
{\eta}_{a}^{A}{\eta}_{b}^{B}{\mathcal{G}}_{\! AB}= \bar{g}_{ab} \label{eq:gab}\vspace{1mm}\\
k_{\mu\nu a}(x,y) & = &
-\eta_{a,\mu}^{A}{\mathcal{Z}}_{,\nu}^{B}{\mathcal{G}}_{AB} \nonumber\vspace{1mm}\\
\hspace{-6mm}\phantom{x}\hspace{-10mm}&=&\bar{k}_{\mu\nu 
a}\!\!-\delta y^{b}\bar{g}^{\alpha\beta}\bar{k}_{\mu\alpha 
a}\bar{k}_{\nu\beta
b}-\!\!{g}^{cd}\delta y^{b}\bar{A}_{\mu ca}\bar{A}_{\nu db},\hspace{3mm}\label{eq:kmunua} \vspace{1mm}\\
A_{\mu ab}(x,y) & = & 
\eta_{a,\mu}^{A}\eta_{b}^{B}{\mathcal{G}}_{AB}\!\!=\!\!\bar{A}_{\mu 
ab}(x)\label{eq:A} \end{eqnarray}
 The contravariant components of the perturbed geometry must be consistent with
${\cal G}^{AC}{\cal G}_{CB} = \delta^A_B$, which can be realized by 
setting $g_{\mu\rho}g^{\rho\nu}= \delta^{\nu}_{\mu}$, $g_{ac}g^{cb}= 
\delta^{b}_{a}$. Since the indices $\mu$ and $b$ can never be equal, 
we must nave $g_{\mu\rho}g^{\rho b}+g_{\mu c}g^{c b}= 
\delta_{\mu}^{b}=0$. After some algebra we see that this corresponds 
to an identity $y^m y^n g^{ab}A_{\mu a m}A_{\nu b n}=-y^b y^n 
g^{ma}A_{\mu [a m]}A_{\nu [b n]} \equiv 0$.

Comparing \rf{gmunu} and \rf{kmunua} we obtain
 \begin{equation}
 k_{\mu\nu
a}=-\frac{1}{2}\frac{\partial g_{\mu\nu}}{\partial y^{a}} 
\label{eq:yorkg} \end{equation}

Consequently, the local bulk defined in a neighborhood around 
$\bar{\sigma}_4$, is foliated by this perturbed geometry, so that 
the Riemann curvature of the bulk may be expressed in the perturbed 
embedding frame. For any fields in the bulk $\xi$ and $\zeta$, the 
covariant derivative $D_\xi \zeta$ is defined by the metric affine 
connection $\Gamma_{ABC}$, with the Riemann tensor given by $ {\cal 
R}(\xi,\zeta) =[D_{\xi},D_{\zeta}]$. Writing the components of this 
tensor in the embedding frame we obtain  the Gauss, Codazzi and 
Ricci equations, respectively:
 \begin{eqnarray}
&&\phantom{x}\hspace{-1,5cm}{\cal R}_{ABCD}{\cal Z}^{A}_{,\alpha} 
{\cal Z}^{B}_{,\beta}{\cal Z}^{C}_{,\gamma}{\cal Z}^{D}_{,\delta} 
=R_{\alpha\beta\gamma\delta} - 2g^{mn}k_{\alpha[\gamma 
m}k_{\delta]\beta n}
 \label{eq:Gauss}\\
&&\phantom{x}\hspace{-1,6cm}{\cal R}_{ABCD} {\cal Z}^{A}_{,\alpha} 
\eta^{B}_{b}{\cal Z}^{C}_{,\gamma}{\cal Z}^{D}_{,\delta} 
=k_{\alpha[\gamma b; \delta]} -
g^{mn}A_{[\gamma mb}k_{\alpha\delta]n }\label{eq:Codazzi}\\
&&\phantom{x}\hspace{-1,8cm}{\cal R}_{ABCD}\eta^{A}_{a}\eta^{B}_{b} 
{\cal Z}^{C}_{,\gamma} {\cal Z}^{D}_{,\delta} = -2g^{mn}A_{[\gamma 
ma}A_{\delta]n b} -2A_{[\gamma a b ; \delta]} - g^{\mu\nu}k_{[\gamma 
\mu a}k_{\delta]\nu b} \label{eq:Ricci} \end{eqnarray} which are the 
integrability conditions for the embedding.
 The differentiable embedding occurs when for a given Riemann tensor for the bulk these equations can be solved without appeal to analyticity. A substantial part of Nash's theorem consists in showing that the solution requires that the functions appearing in the right hand side must be regular.

The expression \rf{yorkg} shows that besides the brane-world 
gravitational field the extrinsic curvature $k_{\mu\nu a}$ also 
propagate in the bulk. The implication of this is that the 
imposition of any restrictive conditions on $k_{\mu\nu a}$ also 
implies on restrictions on the propagation of the gravitational 
field of the brane-world. On the other hand, from \rf{A} it follows 
that the third fundamental form $A_{\mu ab}$ does not propagate at 
all in the bulk, behaving as if it is a confined field.

The equations of motion of the brane-world follow directly from the 
Einstein-Hilbert principle on the bulk and from the integrability 
conditions \rf{Gauss}-\rf{Ricci}. To see how this works take the 
trace of the first equation \rf{Z}:
 $g^{\mu\nu}Z^{A}_{,\mu}Z^{B}_{,\nu}{\cal G}_{AB}= D-N =
{\cal G}_{AB}{\cal G}^{AB}-g^{ab} g_{ab}$, and replace $g_{ab}$ from 
\rf{Z}, obtaining \begin{equation} 
g^{\mu\nu}Z^{A}_{,\mu}Z^{B}_{,\nu} = {\cal G}^{AB} 
-g^{ab}\eta^{A}_{a}\eta^{B}_{b}\label{eq:invert} \end{equation} The 
contractions of \rf{Gauss} with $g^{\mu\nu}$, and using \rf{invert} 
gives the the Ricci tensor and Ricci scalar of the brane-world 
respectively expressed as \begin{eqnarray}
 R_{\mu\nu} & =
&g^{cd}(g^{\alpha\beta}k_{\mu\alpha c} k_{\nu \beta d} - 
h_{c}k_{\mu\nu d}) +{\cal R}_{AB}Z^{A}_{,\mu}Z^{B}_{,\nu} - 
g^{ab}{\cal R}_{ABCD}\eta^{A}_{a} 
Z^{B}_{,\mu}Z^{C}_{,\nu}\eta^{D}_{b}\label{eq:RicciT} \end{eqnarray} 
After another contraction with $g^{\mu\nu}$,
using again \rf{invert}, and noting that\\
$g^{ad}g^{bc}{\cal 
R}_{ABCD}\eta^{A}_{a}\eta^{B}_{b}\eta^{C}_{c}\eta^{D}_{d}=0$, we 
obtain the Ricci scalar \begin{eqnarray} R =(K^{2} -h^{2}) +{\cal R} 
-2g^{ab}{\cal R}_{AB} \eta^{A}_{a}\eta^{B}_{b}
  \label{eq:RICCI}
\end{eqnarray} where $K^{2}=g^{ab}k^{\mu\nu}{}_{a}k_{\mu\nu b}$. 
$h_{a}=g^{\mu\nu}k_{\mu\nu a}$ and $h^{2} =g^{ab }h_{a}h_{b}$. 
Therefore the Einstein-Hilbert action for the bulk geometry in 
$D$-dimensions can be written as \begin{eqnarray} {\cal A}= 
\int{{\cal R}}\sqrt{\cal{G}}d^{D}v =&& \int{\left[ R -( K^{2} - 
h^{2})+
2g^{ab} {\cal R}_{AB}\eta^{A}_{a}\eta^{B}_{b} \right] \sqrt{\cal G}d^{D}v}\nonumber\\
&&\phantom{x}\hspace{4cm} =\alpha_{*}\int {\cal 
L}^{*}\sqrt{\cal{G}}d^{D}v \label{eq:EH} \end{eqnarray} where 
$\alpha_*$ denotes the fundamental energy scale in the bulk and 
${\cal L}^{*}$ is the source Lagrangian. The Euler-Lagrange 
equations of \rf{EH} with respect to ${\cal G}_{AB}$ are Einstein's 
equations in $D$ dimensions: \begin{equation} {\mathcal{R}}_{AB} 
-\frac{1}{2}{\mathcal{R}}{\mathcal{G}}_{AB}= \alpha_{*}T_{AB}^{*} 
\label{eq:bulkEE2} \end{equation} Here $T_{AB}^*$ denote the 
components of the energy-momentum tensor of the sources.

The equations of motion of the embedded brane-world can be derived 
directly from the components of \rf{bulkEE2} written in the 
embedding frame. The tangent components follow from the contractions 
of \rf{bulkEE2} with $\Z^A_{,\mu}\Z^B_{,\nu}$. After using 
\rf{RicciT} and \rf{RICCI} we obtain \begin{eqnarray} 
R_{\mu\nu}-\frac{1}{2}R g_{\mu\nu}- Q_{\mu\nu} -W_{\mu\nu} - 
g^{ab}{\cal R}_{AB}\eta^A_a \eta^B_b =\alpha_* T^*_{\mu\nu} 
\label{eq:BE1} \end{eqnarray} where we have denoted \begin{eqnarray}
 Q_{\mu\nu} & = & g^{ab}k^{\rho}{}_{\mu
a}k_{\rho\nu b}-g^{ab}h_{a}k_{\mu\nu 
b}-\frac{1}{2}(K^{2}-h^{2})g_{\mu\nu}
\label{eq:Qmunu}\\
W_{\mu\nu} & = & 
g^{ad}{\mathcal{R}}_{ABCD}\eta_{a}^{A}{\mathcal{Z}}_{,\mu}^{B} 
{\mathcal{Z}}_{,\nu}^{C}\eta_{d}^{D}\nonumber \end{eqnarray} By a 
direct calculation we can see that the extrinsic tensor $Q_{\mu\nu}$ 
is an independently conserved quantity with respect to the 
brane-world metric.

The contraction of \rf{bulkEE2} with $\Z^A_\mu \eta^B_b$ gives a 
vectorial equation. Using \rf{RICCI} and \rf{Codazzi} we obtain 
\begin{eqnarray}
 &&\phantom{x}\hspace{-14mm}
k_{\mu a;\rho}^{\rho}\! -\!h_{a,\mu} \!+\! A_{\rho c a}k^{\rho 
\;c}_{\;\mu}\! -\!A_{\mu c a}h^{c}
 \! + 2W_{\mu a}= -2\alpha_* (T^*_{\mu a} -\frac{1}{N+2}T^* g_{\mu a})\label{eq:BE2}
\end{eqnarray} where we have denoted \begin{equation} W_{\mu a}= 
g^{bd} {\cal R}_{ABCD}\eta^{A}_{a}\eta^{B}_{b}{\cal 
Z}^{C}_{,\mu}\eta^{D}_{d} \nonumber \end{equation} Finally, 
contracting \rf{bulkEE2} with $\eta^A_a \eta^B_b$ we obtain 
$N(N+1)/2$ scalar equations involving the so called Hawking-Gibbons 
term $S_{ab} = {\cal R}_{AB}\eta^A_a\eta^B_b$ and its trace 
$S=g^{ab}S_{ab}$
 \begin{equation}
S_{ab} -Sg_{ab}- \frac{1}{2}[R-K^{2} +h^{2}]g_{ab} =\alpha_* 
T^*_{ab} \label{eq:BE3} \end{equation}
  In its most general form, without assuming extra dimensional matter, the confinement hypothesis states that the only non-vanishing components of $T_{AB}$ are the tangent components $T_{\mu\nu}$ representing the confined sources \cite{ADD}. Therefore we set
\begin{eqnarray} &&\alpha_{*} T^{*}_{AB}{\cal Z}^{A}_{,\mu}{\cal
Z}^{B}_{,\nu}= \alpha_{*} T_{\mu\nu}^{*} =-8\pi G T_{\mu\nu}\\
&&\alpha_* T^{*}_{AB}{\cal Z}^{A}_{,\mu}\eta^{B}_{a} =\alpha_{*} T_{\mu a}^{*}=0\\
&&\alpha_{*} T^{*}_{AB}\eta^{A}_{a}\eta^{B}_{b}\; =\alpha_* 
T_{ab}^{*}= 0 \end{eqnarray} Equations \rf{BE1}, \rf{BE2} and 
\rf{BE3} with confinement conditions are sometimes called the 
gravi-tensor, gravi-vector and gravi-scalars (Usually a single 
gravi-scalar equation in the 5-dimensional models \cite{Durrer}) 
equations respectively. These represent generalizations of 
Einstein's equations of general relativity, in the sense that they 
describe the evolution of all geometrical components $g_{\mu\nu}$, 
$A_{\mu ab}$ and $k_{\mu\nu a}$ of the brane-world. Clearly, the 
usual Einstein's equations are recovered when all elements of the 
extrinsic geometry are removed from those equations.

\section{Canonical Equations} The standard ADM canonical 
quantization of the gravitational field in general relativity was 
originally intended to describe the quantum fluctuations of 
3-dimensional hypersurfaces in a space-time \cite{ADM}. The 
space-time metric is decomposed in 3-surface components, plus a 
shift vector and a lapse function defined in a Gaussian reference 
frame defined on the 3-dimensional hypersurface. After writing the 
Einstein-Hilbert Lagrangian in this Gaussian frame, the 
Euler-Lagrange equations with respect to the shift leads to the 
vanishing of the Hamiltonian. This is not a real problem because in 
principle the system could be solved by use of Dirac's standard 
procedure for constrained systems. However, as it is well known, the 
Poisson bracket structure does not propagate covariantly as it would 
be expected. In spite of all efforts made up to the present, this 
problem remains unsolved \cite{Kuchar,Isham,Hojman,Alvarez}. It is 
possible to describe a non-constrained canonical system in a special 
frame defined by a 3-dimensional hypersurface orthogonal Gaussian 
coordinate system. In such special frame the shift vector vanishes 
and the Hamiltonian constraint does not apply \cite{Dirac}. 
Nonetheless, this has been regarded as of little value for general 
relativity itself, essentially because the diffeomorphism group of 
the theory is one of the fundamental postulates of the theory 
\cite{Diracham}.

 The extension of the ADM canonical formulation to the brane-world is straightforward but it requires a few adaptations: First, the bulk is locally foliated by a continuous sequence of brane-worlds propagating along the extra dimensions rather than by a 3-surface propagating along a single time direction. Secondly, the confinement hypothesis implies that the diffeomorphism invariance do not extend to the extra dimensions, otherwise a coordinate transformation in the bulk would have the effect of introducing a component of the energy-momentum tensor of the confined fields and ordinary matter in the bulk. Therefore, in order to maintain the intended solution of the hierarchy problem, the diffeomorphism of the brane-world must be restricted as a confined symmetry. Actually this can be regarded as one of the merits of brane-world theory, which differentiates it from being just a higher dimensional version of general relativity.
However, to deserve such merit the extra dimensions need to be taken 
seriously as true physical degrees of freedom in the canonical 
formulation of the theory. The momentum conjugated to the metric 
field ${\cal G}_{AB}$, with respect to the displacement along 
$\eta_a$ is defined as usual \[
 p^{AB}{}_{ (a)} =\frac{\partial
{\cal L}}{ \partial \left( \frac{ \partial {\cal G}_{AB} }{\partial 
y^{a}} \right) } \] where $\cal L$ is the Einstein-Hilbert 
Lagrangian of the bulk in \rf{EH}. Noting that in the embedding 
frame we can write $ 2g^{ab}{\cal R}_{AB}\eta^A_a\eta^B_b = K^2 
-g^{ab}h_{a,b} $, after eliminating the divergence term 
$g^{ab}h_{a,b}$, the Lagrangian can be simplified to \be {\cal L} = 
[R + (K^2 + h^2)]\sqrt{\cal G} \label{eq:Lagrangian} \ee Therefore, 
using \rf{yorkg} we obtain the canonically conjugated momenta 
\begin{eqnarray} && p^{\mu\nu}{}_{(a)}=\frac{\partial {\cal L}}{ 
\left(\frac{ \partial g_{\mu \nu} }{\partial y^{a}} 
\right)}=-\frac{1}{2}\frac{\partial {\cal L}}{ \partial k_{\mu\nu a} 
} =-(k^{\mu\nu}{}_{a}
+h_{a}g^{\mu\nu})\sqrt{{\cal G}} \label{eq:MOMENTUM}\\
&& p^{\mu a}{}_{(b)}=\frac{\partial {\cal L}}{ \left( \frac{ 
\partial g_{\mu a} }{\partial y^{b}} \right)}= \frac{\partial {\cal 
L}}{ \partial A_{\mu a b} }=0,\;\;\;
p^{a b}{}_{(c)}=\frac{\partial {\cal L}}{ \left( \frac{ \partial 
g_{a b} }{\partial y^{c}} \right)}=0
\end{eqnarray} The last two components are equal to zero because the 
Lagrangian \rf{Lagrangian} does not depend explicitly on $A_{\mu 
ab}$ and on $g_{ab,c}$.

Using the above expressions, the Hamiltonian ${\cal H}_a$ 
corresponding to the displacement along each orthogonal direction 
$\eta_{a}$ separately, follows from a partial Legendre 
transformation (no sum on (a)): \begin{eqnarray*} &&{\cal 
H}_{a}(g,p_{(a)}) = p^{AB}{}_{(a)}g_{AB,a}-{\cal L}
 =
-R\sqrt{\cal G}-[(K^2 +h^2 +2(K_a^2+h_a^2)]\sqrt{\cal G} 
\end{eqnarray*} where we have denoted $K^2_a= k^{\mu\nu}_a k_{\mu\nu 
a}$, $K^2 =g^{ab}K_a K_b$, and $p_{(a)} =g^{\mu\nu} p_{\mu\nu (a)}$. 
After replacing $h_a =\frac{-p_{(a)}}{5\sqrt{\cal G}}$ it follows 
that
 \be 
{\cal H}_{a} = -R\sqrt{\cal G}-\frac{1}{\sqrt{\cal 
G}}\left[\frac{3p^2}{5} + 2\frac{ p_{(a)}^2}{5} +
 p^{\mu\nu}_{(a)}p_{\mu\nu (a)} 
\right]\label{eq:HA}
\ee For a given functional ${\cal F}(g_{\mu\nu},p_{\mu\nu})$ defined 
in the phase space of the brane-world, the propagation of ${\cal F}$ 
along $y^a$ is given by the Poisson brackets with each Hamiltonian 
separately \be \frac{\tilde\delta {\cal F}}{\tilde\delta 
y^{a}}=[{\cal F},{\cal H}_{a}]=\frac{\tilde\delta {\cal 
F}}{\tilde\delta g_{\mu\nu}} \frac{\tilde\delta {\cal 
H}_{a}}{\tilde\delta p^{\mu\nu(a)}}-\frac{\tilde\delta {\cal 
F}}{\tilde\delta p^{\mu\nu (a)}}\frac{\tilde\delta {\cal 
H}_{a}}{\tilde\delta g_{\mu\nu}} \label{eq:poisson}
 \ee
Here $\tilde\delta$ denotes the standard functional derivative in 
phase space.

 Hamilton's equations for the brane-world with respect to each extra coordinate $y^{a}$ may be written as
\begin{eqnarray} \frac{d g_{\mu\nu}}{d y^{a}} = \frac{\tilde\delta 
{\cal H}_{a}}{\tilde\delta p^{\mu\nu (a)}}=& &
[g_{\mu\nu}, {\cal H}_a ]=-2k_{\mu\nu a}, \label{eq:DOTG}
\phantom{x}\vspace{3mm}\\
&&\phantom{x}\hspace{-3.3cm}\frac{d p^{\mu\nu}_{(a)} }{ d y^{a} }= -\frac{\tilde\delta {\cal H}_{a}}{\tilde\delta g_{\mu\nu}}= 
[p^{\mu\nu}_{(a)}, {\cal H}_a]\label{eq:DOTP} \end{eqnarray} As it 
can be seen, the differences between the brane-world canonical 
formulation and the ADM formulation of general relativity follow 
from the non-vanishing of the Hamiltonians ${\cal H}_a$, as a 
consequence of the brane-world scheme for solving the hierarchy 
problem. With this result, the ADM quantization program can be 
retaken, with the difference that the quantum equation should 
describe the "states" of four-dimensional submanifolds in the bulk, 
with respect to the extra dimensions.

\section{Tomonaga-Schwinger Quantum States} The Tomonaga-Schwinger 
equation originated from Dirac's many fingered time formalism for 
relativistic quantum theory, in which a set of $N$ electrons was 
associated to $N$ proper times satisfying $N$ Schrodinger's-like 
equations \cite{Diracfingers}. The continuous limit of this equation 
was formulated by Tomonaga for a relativistic field defined in a 
region of space-time characterized by an evolving space-like 
3-hypersurface $\sigma$ with a time direction attached to each of 
its point. This geometric extension of Dirac's many fingered time, 
which was soon realized to be equivalent to the interaction 
representation of quantum mechanics developed by Schwinger 
\cite{Tomonaga,Schwinger,Nishijima}. Here, it is more convenient to 
look at the Tomonaga-Schwinger equation from the geometrical point 
of view written as
 \be
i\hbar\frac{\delta \Psi}{\delta \sigma} = \hat{\cal H}_{\sigma}\Psi 
\label{eq:TS} \ee which represents a generalization of Schrodinger's 
equation, describing the quantum state functional $\Psi$ of a 
space-like 3-hypersurface $\sigma$ embedded in Minkowski space-time. 
In the right hand side, the Hamiltonian operator describes the 
translational operator along a time-like direction orthogonal to 
$\sigma$. The functional derivative in the left hand side is defined 
by the limit \be \frac{\delta \Psi }{\delta \sigma} = \lim_{\Delta 
V\rightarrow 0}\frac{\Psi ( \sigma')-\Psi(\sigma)}{\Delta V} 
\label{eq:delta} \ee where $\Delta V$ denotes the local volume 
element between two neighboring hypersurfaces $\sigma$ and 
$\sigma'$.

The main difficulty of \rf{TS} is that it is not easily integrable. 
In the particular case where $[\hat{\cal H}_\sigma,\hat{\cal 
H}_{\sigma'}]=0$, the equation can be integrated, but the 
hypersurfaces $\sigma$ and $\sigma'$ are necessarily flat. In the 
general case where the hypersurfaces are not flat, the solutions of 
\rf{TS} can be determined as an approximate solution after the 
application of the Yang-Feldmann formalism and Dyson's expression 
for the $S$ matrix \cite{Nishijima,Pauli}. The difficulty in solving 
\rf{TS} can be traced back to the fact that the limit operation in 
\rf{delta} was not defined. In fact, the conditions to decide how 
close $\sigma$ and $\sigma'$ are were not given previously, and it 
can be decided only after solving the quantum equation itself using 
some quantum approximation method.

In the application of \rf{TS} to the brane-world, the limit 
operation between two four-dimensional brane-worlds $\sigma_4$ and 
$\sigma'_4$ is improved because Nash's theorem shows at the 
classical level how to tell the separation between the two sub 
manifolds. In other words, since each brane-world was generated by 
classical perturbations of an initial sub manifold $\sigma_4$, the 
volume element in \rf{delta} has been already specified by the 
parameter $\delta y^a$ of the perturbed geometry. In practice, we 
may split the bulk volume $\Delta V$ between $\sigma_4$ and 
$\sigma'_4$ into a product of the volume $\Delta v$ of a a small 
compact region in $\sigma_4$, times the variation $\Delta y^a$ of 
the extra dimensional coordinates $y^a$. Therefore, it sufficient to 
specify only the limit operation $\Delta y^a \rightarrow 0$ and the 
functional derivative \rf{delta} for density functions with compact 
support on the brane-world, with respect to each extra dimension can 
be simplified to $$ \left. \frac{\delta \Psi }{\delta 
\sigma}\right\rfloor_{y^a} = \lim_{\Delta y^a\rightarrow 
0}\frac{\Psi ( \sigma')-\Psi(\sigma)}{\Delta y^a} =\frac{\partial 
\Psi}{\partial y^a} $$ Repeating for all extra dimensions, we find 
that the Tomonaga-Schwinger equation \rf{TS} can be extended the 
brane-world, as a system of $N$ partial equations, one for each 
extra dimension
 \be
i\hbar\frac{\partial \Psi_a}{\partial y^a} = \hat{\cal H}_a\Psi_a, 
\;\;\; a = 5..D \label{eq:TSBW} \ee which gives to $\hat{\cal H}_a$ 
the interpretation of the extra dimensional translational operator. 
The final quantum state is given by the superposition of the $N$ 
separates states $\Psi_a$ as $\Psi = \sum B^a \Psi_a$. The state 
functional density $\Psi$ represents the quantum fluctuation of the 
brane-world sub manifold in the bulk at the (Tev) energy scale, 
subjected to quantum uncertainties and a state probability given by 
$$ || \Psi ||^2 = \int \Psi^\dagger \Psi \delta y \delta v $$ An 
observer confined to the brane-world may evaluate the quantum 
expectation values of the brane-world metric and the extrinsic 
curvature are given by $$ <\Psi| \hat{g}_{\mu\nu}|\Psi > =\int 
\Psi^\dagger \hat{g}_{\mu\nu}\Psi \delta y \delta v,\; \;\;\;\; 
<\Psi| \hat{k}_{\mu\nu}|\Psi > =\int \Psi^\dagger 
\hat{k}_{\mu\nu}\Psi \delta y \delta v $$ Since the classical 
$k_{\mu\nu a}$ is the derivative of the metric $g_{\mu\nu}$ with 
respect to $y_a$, we may set boundary conditions on these quantities 
at the initial brane-world $y^a =0$ to determine the final solution.

\section{Overview and Perspectives}

We have shown that the Einstein-Hilbert principle applied to the 
bulk geometry plus the differentiable embedding conditions are 
sufficient to determine the classical and quantum structures of the 
brane-world in D-dimensions. In particular, it was shown that Nash's 
theorem makes it possible to generate any embedded sub manifold by a 
continuous sequence of infinitesimal perturbations of an arbitrarily 
given embedded geometry along the extra dimensions. Using the 
classical perturbative embeddings, and the basic principles of the 
brane-world theory we have obtained a canonical structure very much 
like the ADM canonical formulation of general relativity, with the 
exception that the Hamiltonians do not vanish.

 The definition of the functional derivatives was improved with respect to four-dimensional field theory, by using the previously defined perturbative embedding structure of the brane-world. The quantization of the brane-world was described the Tomonaga-Schwinger equation defined for brane-world sub manifolds, calculated for each extra dimension. Actually, as a result of the the the classical perturbation theory, the Tomonaga-Schwinger equation becomes exactly integrable.
\vspace{2mm}

 In view of current astrophysical observations and the near future high energy experiments, there are some applications of the quantum brane-world theory to be detailed in subsequent papers:
\vspace{3mm}

(a)\underline{ Brane-world Cosmology}\\
Brane-world cosmology offers a possible explanation to the 
accelerated expansion of the universe, resulting from the 
modification of the Friedman's equation by the presence of the 
extrinsic curvature included in the tensor $Q_{\mu\nu}$ given by 
\rf{Qmunu} \cite{MaiaGDE}. The presence of this tensor has the 
meaning that the brane-world vacuum is more complex than the vacuum 
in general relativity. In fact, for a constant curvature bulk with 
curvature $\Lambda_*$, after eliminating redundant terms, the 
gravi-tensor vacuum equation becomes $R_{\mu\nu}- 1/2 R g_{\mu\nu} 
-Q_{\mu\nu} + \Lambda_* g_{\mu\nu}= 0$.

 Therefore, the vacuum energy density $<\rho_v>$ resulting from gravitationally coupled fields must be revaluated, including the extrinsic curvature component. This suggests that in some epochs, say at the early inflationary period, the extrinsic curvature may contribute to the vacuum energy, differently from other periods.

The particular case where we have only one extra dimensions ($D=5$), 
has some limitations with respect to the differentiable embedding. 
However, some cosmological models like the FRW, deSitter and anti 
deSitter solutions of Einstein's equations in four dimensions can be 
embedded in five dimensional bulks without restrictions, in 
accordance with the perturbative embedding equations previously 
shown. Consequently, in such brane-world cosmological models the 
conditions required for a proper definition of the functional 
derivatives in the Tomonaga-Schwinger equation are well established, 
and the equation can can be integrated without difficulty.

(b) \underline{Laboratory production of mini black holes}:\\
Brane-world gravity predict the generation of short lived mini black 
holes produced at the Tev energy in the laboratory, resulting from 
proton-proton collisions \cite{Dimopolous}. However, using 
semi-classical quantum gravity in four dimensions, we have learned 
that quantum unitarity does not necessarily hold true during the 
black hole evaporation. On the other hand, using Euclidean path 
integral, it was shown that the unitarity can be restored with the 
aid of the ADS/CFT correspondence in the framework of $AdS_5 \times 
S^5$ string theory \cite{Hawking}.

 Since the generation of mini black holes are possible only in the brane-world context, the whole process includes the original Minkowski's space-time where the experiment is devised. Soon after the collision, the space-time must be deformable into a Schwarzschild or a Reissner-Nordstrom black hole. Finally, after a short period of evaporation the space-time may be back to Minkowski's configuration or else leaves a curved remnant. The description of such process can start with the classical perturbations in accordance with Nash's embedded geometries, but the unitarity is has to be decided at the quantum level. In this respect we notice that both Schwarzschild and Reissner-Nordstrom black holes are well defined submanifolds embedded in a six-dimensional flat bulk with signature $(4,2)$. In this case, the bulk isometry group $SO(4,2)$ is isomorphic to the conformal group in Minkowski's space-time, compatible with the ADS/CFT correspondence adapted to the brane-world \cite{Maiaessay}. Therefore, the quantum unitarity implicitly assumed in the Tomonaga-Schwinger equation, must be consistent with the black hole evaporation theorems in six dimensions.

(c) \underline{Quantum Four-manifold Theory}:\\
The above description of quantum theory of the brane-world is based 
almost entirely on the general theory of differentiable sub 
manifolds. This suggests a quantum theory of four-dimensional sub 
manifolds. It starts with the classical perturbations of embedded 
geometries, but ends with a quantum version of the embedding 
theorem, including the fluctuations of the embedding as described by 
the Tomonaga-Schwinger equation. This quantum theory of submanifolds 
would be particularly interesting when the bulk has dimensions 
greater than five, where the third fundamental form behaves 
similarly to a gauge field with respect to the extra dimensional 
group of isometries. The identification of the third fundamental 
form as a gauge field with the symmetry of the extra dimensions 
plying the role of the gauge group is old, but it was never taken 
seriously \cite{Ne'emann}. \vspace{2mm}

One frequent criticism to string theory is that it depends on a 
pre-existing background space-time with 10 (or 11) dimensions, 
acting as the host space for all possible dynamics 
\cite{Ashtekar,Thiemann}. On the other hand, loop quantum gravity 
does not require such background, but it depends on a previously 
existing spin network structure \cite{Nicolai}. Quantum brane-world 
gravity does not have a background space in the same sense of string 
theory because the bulk is the primary dynamical object. The 
Einstein-Hilbert principle applied to the bulk geometry provides all 
dynamics of the brane-world, without requiring any new algebraic 
structure besides the theory of differentiable manifolds, where our 
basic notions of space, topology and analysis begin and make sense.

\end{document}